\documentclass[12pt,a4paper,final]{iopart}
\usepackage{iopams}
\usepackage[usenames,dvipsnames]{xcolor}
\usepackage[latin9]{inputenc}
\usepackage{graphicx}
\usepackage{esint}
\usepackage{comment}
\usepackage{tensor}
\usepackage{ulem}
\usepackage{mathrsfs}
\usepackage{amsthm}
\usepackage{bm}
\everymath{\displaystyle}
\usepackage{hyperref}
\definecolor{MyDarkRed}{rgb}{0.71,0.14,0.07}
\definecolor{MyDarkBlue}{rgb}{0.1,0,0.7}
\definecolor{MyDarkGreen}{rgb}{0.11,0.64,0.22}
  \hypersetup{
    colorlinks,%
    citecolor=blue,%
    filecolor=blue,%
    linkcolor=magenta,%
    urlcolor=blue
}

\newcommand\eprintarXiv[1]{\href{http://arXiv.org/abs/#1}{#1}}
\newcommand{\ddt}[1]{\frac{{\mathrm d} #1}{{\mathrm d} t}}
\newcommand{\beq}{\begin{equation}}
\newcommand{\eeq}{\end{equation}}
\newcommand{\beqNo}{\begin{equation*}}
\newcommand{\eeqNo}{\end{equation*}}

\newcommand{\cc}{\mathrm{\Lambda}}
\newcommand{\CA}{\mathcal A}

\newcommand{\CD}{\mathcal D}

\newcommand{\CL}{\mathcal L}

\newcommand{\CP}{\mathcal P}
\newcommand{\CQ}{\mathcal Q}

\newcommand{\CW}{\mathcal W}

\newcommand{\QD}{\CQ_\CD}
\newcommand{\PD}{\CP_\CD}

\newcommand{\SR}{\mathscr R}

\newcommand{\WD}{\mathcal W_{\mathcal D}}
 1

\newcommand{\epsilont}{\tilde \epsilon}
\newcommand{\omegat}{\tilde \omega}

\newcommand{\sigmat}{\tilde \sigma}
\newcommand{\SRt}{\tilde {\mathscr R}}
\newcommand{\Thetat}{\tilde \Theta}
\newcommand{\baverage}[1]{\left\langle {\strut #1}\right\rangle_{\CD}}

\makeatletter
\long\def\@makefntext#1{\parindent 1em\noindent 
 \makebox[1em][l]{\footnotesize\rm$\m@th{^{\arabic{footnote}}}$}%
 \footnotesize\rm #1}
\def\@makefnmark{\hbox{$^{\arabic{footnote}}\m@th$}}
\def\@thefnmark{\arabic{footnote}}
\makeatother
\begin{document}
\letter{Cosmological backreaction and its dependence on spacetime foliation}
\author{Thomas Buchert, Pierre Mourier and Xavier Roy}
\address{Univ Lyon, Ens de Lyon, Univ Lyon1, CNRS, \\
Centre de Recherche Astrophysique de Lyon UMR5574, F--69007 Lyon, France\\
\eads{\mailto{buchert@ens-lyon.fr}, \mailto{pierre.mourier@ens-lyon.fr}, \mailto{x.roy@gmx.com}}
}
\begin{abstract}
      The subject of cosmological backreaction in General Relativity is often approached by coordinate-dependent and 
      metric-based analyses. We present in this letter an averaging formalism for the scalar parts of Einstein's equations that is coordinate-independent and only functionally depends on a metric. This formalism is applicable to general $3+1$ foliations of spacetime for an arbitrary fluid with tilted flow. We clarify the dependence on spacetime foliation and argue that this dependence is weak in cosmological settings. We also introduce a new set of averaged equations that feature a global cosmological time despite the generality of the setting, and we put the statistical nature of effective cosmologies into perspective.
\end{abstract}
\section{Context}
\label{context}
Cosmology deals with models for the evolution of the Universe and, within General Relativity, entails the question of how to split the $4-$dimensional spacetime into a $3-$dimensional space evolving in time. This question can be formally answered by a $1+3$ threading along a preferred timelike congruence (see, e.g. \cite{Ellisbook,Jantzen}), or by a $3+1$ slicing (foliation) into a family of spacelike hypersurfaces (see, e.g. \cite{adm, gourg:foliation}). 
Both decompositions introduce four degrees of freedom, which are given in terms of a lapse function and a shift vector (or one-form). We shall consider the threading and slicing approaches jointly to formalize configurations where the fluid content is described by a $4-$velocity tilted with respect to the hypersurface normal.
{\it A priori,} only in special cases does the slicing keep the proper time of the fluid elements synchronous.

In standard cosmology one commonly idealizes the 
geometry of the Universe by a homogeneous-isotropic background metric with constant spatial curvature. In the case of the so-called concordance or $\Lambda$CDM model (``Cold Dark Matter with dark energy modeled by the cosmological constant $\Lambda$''), the metric form features a global time $t$ labeling Euclidean spatial sections that admit global coordinates $x^i$, with a global rescaling factor $a(t)$,\footnote[1]{We adopt the conventions that 
Greek indices are assigned to spacetime indices running in $\{ 0, 1, 2, 3 \}$, and Latin indices refer to space indices, running in $\{ 1, 2, 3 \}$. The signature of the metric is taken as $(- + + +)$, and the units are such that $c = 1$.} ${}^{4}{\mathbf g}^{\mathrm hom} = - {\mathbf d}t^2 + a^2 (t) \delta_{ij} {\mathbf d}x^i \otimes {\mathbf d}x^j$. 
It is known that other choices of slicing change Friedmann's equations which determine the only gravitational degree of freedom, $a(t)$. See, e.g. \cite{buchert:av_pf}. The scale factor may even become space-dependent, for instance for a general slicing lapse. The problem of dependence on spacetime foliation therefore also exists in standard cosmology, where it is solved by choosing a preferred (proper time) foliation anchored to the Cosmic Microwave Background (CMB) rest frame.

Nonlinear structure formation in cosmology is most often investigated within the Newtonian framework of self-gravitating fluids.
Efforts to describe relativistic models of inhomogeneities often rely on metric forms that are designed to be `close to'
the homogeneous-isotropic metric form above. The idea is to describe `small' perturbations, which is a sensible assumption for metric perturbations, since they are indeed very small except in the vicinity of strong field astrophysical objects \cite{ishibashiwald}.\footnote[2]{\label{footnote2}However, the derivatives of the metric can be large. Even for metric perturbations of order $10^{-6}$, curvature perturbations can be of order unity and therefore out of reach in this setting \cite{estimate}. Green and Wald \cite{gw} have modified earlier statements of \cite{ishibashiwald} emphasizing that curvature can be large. (Their statement of trace-free backreaction, however, has no physical justification \cite{buchert11}.)}
In the \textit{longitudinal gauge}, the lapse function and the
spatial $3-$metric are referred to a `perturbed Newtonian setting', with an assumed metric form for the physical spacetime,
\beq
\label{quasinewton1}
{}^{4}{\mathbf g}^{\mathrm pert} = -\,N^{2}{\mathbf d}t^{2} + g_{ij}\,{\mathbf d}x^{i} \otimes {\mathbf d}x^{j} \;,
\eeq
where the slicing lapse $N$ and the $3-$metric coefficients
$g_{ij}$ of a family of spacelike hypersurfaces $t=\mbox{\textit{const.}}$ are written as (here for scalar perturbations only):\footnote[3]{Here, $a(t)$ denotes the same scale factor as in the homogeneous-isotropic case, which follows by setting $0 = \Phi = \Psi$; $\gamma_{ij}$ denotes a constant curvature metric that is commonly considered to be flat, $\gamma_{ij} = \delta_{ij}$.}
\beq
\label{quasinewton2}
N^{2} \equiv 1+2\Phi(t,x^{i}) \quad ; \quad g_{ij} \equiv a^{2}(t)[1-2\Psi(t,x^{i})]\,\gamma_{ij} \, .
\eeq
We note that the extrinsic curvature has no trace-free part,
i.e. for a fluid $4-$velocity parallel to the normal congruence $N^{-1}\,\boldsymbol{\partial}_{t}$ the above metric describes homogeneous solutions in cosmologically relevant cases \cite{col1988,colwai1983,collinswhite2}.\footnote[4]{It is commonly assumed that the $4-$velocity is tilted with respect to the normal congruence, but that spatial velocities are non-relativistic, i.e. that the Lorentz factor $\gamma$ is close to $1$. 
Our remark implies that by replacing the approximate sign by an equality sign the fluid has to be \textit{shear-free} 
in the metric form $\lbrace$(\ref{quasinewton1}) and (\ref{quasinewton2})$\rbrace$ and, hence, \textit{homogeneous} in cosmologically relevant cases \cite{col1988,colwai1983,collinswhite2}. }

Metric forms that are designed to stay `close to' a homogeneous solution are also used to address the backreaction problem by devising simulations that include relativistic corrections. As an example we read in \cite{Adamek2017} (see also references therein)
that `the backreaction from structure can differ by many orders of magnitude depending upon the slicing of spacetime one chooses to average over'. We shall confront this statement with a covariant and background-free result about averaged dynamics that allows us to discuss the foliation dependence of backreaction without the need to consider gauge transformations.

We finally emphasize that cosmological backreaction can only be present if the average spatial curvature, and hence the large-scale average of cosmological variables, are allowed to evolve. Schemes that suppress average curvature evolution by, e.g. employing periodic boundary conditions as in Newtonian models \cite{newtonianbackreaction}, cannot describe global backreaction, but only backreaction in the interior of an assumed background model, i.e. `cosmic variance'.\footnote[5]{Theoretical foundations of the cosmological backreaction effect via structure-emerging average spatial curvature may be found in \cite{buchert:av_dust,buchert:review}. (See also illustrations within a class of background-free simulations in \cite{bolejko:emerging}.)}

\section{Explicit foliation dependence of backreaction}
\label{foliationdependence}

Cosmological backreaction is the study of inhomogeneity effects on the global evolution of the model universe.
This involves averaging strategies which can for instance be unambiguously defined on the basis of volume averages of scalars. For \textit{irrotational dust} and \textit{irrotational perfect fluids} the answer has been given in terms of volume averaged scalar parts of the Einstein equations in \cite{buchert:av_dust, buchert:av_pf, buchert:review}. This yielded cosmological equations of Friedmannian form for an effective energy-momentum tensor including averages of (extrinsic and intrinsic) curvature invariants of geometrical inhomogeneities in fluid-orthogonal spatial domains. These results are background-free, they depend on the averaging domain (e.g. on spatial scale), and they imply a dependence on the metric only via the morphology of the domain and the volume element of integration. As we shall discuss, this implicit dependence on the metric can be exploited for a statistical interpretation of the effective cosmological equations.

In a forthcoming investigation we derive the scalar-averaged equations for arbitrary $3+1$ foliations with general tilted fluid flow \cite{av_general}. There, we discuss in detail relations to other works where such generalizations are offered. These earlier proposals focus on an \textit{extrinsic approach}, i.e. they perform averages of the geometrical variables as seen by hypersurface observers. As we also discuss in \cite{av_general}, this approach inherits problems such as the non-conservation of the number of fluid elements within the averaging domain as it evolves.

We present in this letter the general scalar-averaged equations derived from an \textit{intrinsic approach}, therefore following the spirit of the original works \cite{buchert:av_dust, buchert:av_pf}. Specifically, we 
perform averages of the fluid variables as seen by fluid observers. We consider an arbitrary spatial foliation which can be tilted with respect to the fluid congruence; this is necessary for a general flow as a fluid-orthogonal foliation is impossible as soon as the fluid has nonzero vorticity \cite{ehlers}. Accordingly, 
{\it local} spacelike projections can be performed onto the local tangent spaces of the hypersurfaces of the foliation along their normal $\bm n$, with $h_{\mu\nu} = g_{\mu\nu}+n_{\mu}n_{\nu}$, or onto the rest frames of the fluid elements along their $4-$velocity $\bm u$, with $b_{\mu\nu} = g_{\mu\nu}+u_{\mu}u_{\nu}$. 
These projectors define two covariant volume measures on the tangent spaces of the hypersurfaces: ${\textstyle\sqrt{\det(h_{ij})} \mathrm{d}^3 x}$ and ${\textstyle \sqrt{\det(b_{ij})} \mathrm{d}^3 x = \gamma \sqrt{\det(h_{ij})} \mathrm{d}^3 x}$, with $x^i$ arbitrary local spatial coordinates, and $\gamma$ the Lorentz factor given by the fluid spatial velocity $\bm v$, as a measure of the local tilt between $\bm n$ and $\bm u$, as follows:
\beq
\gamma =  \frac{1}{\sqrt{1 - v^{\alpha} v_{\alpha}}} \ \ ; \ \ u^{\mu} = \gamma (n^{\mu} + v^{\mu}) \ \  ; \ \  n^{\alpha}v_{\alpha} = 0\;.
\label{lorentz}
\eeq
We associate accordingly to the same averaging domain $\CD$ lying in the hypersurfaces two different volumes: the Riemannian volume $V^h_\CD \equiv {\textstyle \int_\CD \sqrt{\det (h_{ij})}}\, {\mathrm d}^3 x$, 
and the fluid \textit{proper volume}, $V_\CD^b \equiv \textstyle{\int_\CD \sqrt{\det (b_{ij})}} \,{\mathrm d}^3 x$. 
The former appears on average Lorentz-contracted with respect to the latter: introducing the \textit{proper~volume averager}, defined for any scalar $\varphi$ as $\langle \varphi \rangle_\CD^b \equiv \textstyle{1/V_\CD^b \int_\CD \varphi \sqrt{\det (b_{ij})}} \, {\mathrm d}^3 x$, we have $V_\CD^h = V_\CD^b \langle 1/ \gamma \rangle^b$, which shows identity in the absence of tilt, i.e. when $\gamma = 1$.
The integral is here again performed over a domain lying within the hypersurfaces of normal $\bm n$. As we shall only consider proper volume averages in the following, we shall omit the index $b$ for notational ease. 

We apply the averaging operator to the scalar parts of the Einstein equations over a compact domain $\CD$ lying within the hypersurfaces. Following \cite{buchert:av_dust, buchert:av_pf}, $\CD$ is chosen to be a {\it comoving domain}, i.e. it is transported along the fluid congruence, which ensures the absence of matter flow across its boundaries and the preservation of its total rest mass. From this procedure we obtain the following \textit{expansion} and \textit{acceleration laws}, together with their \textit{integrability condition}, for rescaled
kinematic fluid variables (the squared rates of expansion, shear and vorticity, $\Thetat^2 = M^2 \Theta^2$, $\sigmat^2 \equiv M^2 \sigma^2$, $\omegat^2 \equiv M^2 \omega^2$), energy density and pressure ($\epsilont \equiv M^2 \epsilon$, $\tilde p \equiv M^2 p$), divergence of the fluid's $4-$acceleration $a^\mu$ ($\tilde{\CA} \equiv M^2 \CA$, with $\CA \equiv \nabla_\mu a^\mu$), and fluid $3-$curvature ($\SRt \equiv M^2 \SR$):\footnote[6]{%
We defined $M \equiv N/\gamma$ (the \textit{threading lapse} in a $1+3$ threading of spacetime). The hypersurfaces are parametrized by a monotonic scalar function $t$. From it we can define the \textit{comoving time-derivative} ${\mathrm d}/{\mathrm d}t$ as the derivative with respect to $t$ along the fluid flow lines, and the \textit{effective Hubble rate}
$H_\CD \equiv ({\mathrm d}a_\CD / {\mathrm d}t ) / a_\CD$ for the \textit{volume scale factor} $a_\CD \equiv \textstyle{(V_\CD / V_{\CD{\mathrm i}})^{1/3}}$. $\Theta$, $\sigma_{\mu\nu}$ and $\omega_{\mu\nu}$ are, respectively, the trace, the symmetric traceless part, and the antisymmetric part of the projected $4-$velocity gradient, $b^{\alpha}_{\ \mu} b^{\beta}_{\ \nu} \nabla_{\alpha} u_{\beta}$. $\sigma^2 : =(1/2) \sigma_{\mu\nu}\sigma^{\mu\nu}$ and
$\omega^2 : =(1/2) \omega_{\mu\nu}\omega^{\mu\nu}$ define the rates of shear and vorticity. The `fluid $3-$curvature' $\mathscr{R}$ is defined from the energy constraint in the fluid rest frames, $\SR \equiv -(2/3) \Theta^2 + 2 \sigma^2 - 2 \omega^2 + 16 \pi G \epsilon + 2 \Lambda$ (see \cite{ellis:vorticity}), and reduces to the $3-$Ricci scalar of the 
fluid-orthogonal hypersurfaces for vanishing vorticity.}%
\begin{eqnarray}
	& 3 \, \frac{1}{a_\CD} \frac{{\mathrm d}^2 a_\CD}{{\mathrm d}t^2} = 
		 \; - 4 \pi G \baverage{ \, \epsilont + 3 \tilde p \, } + {\tilde\Lambda}_\CD + \tilde{\CQ}_\CD
			+ \tilde{\CP}_\CD \; ;  
\nonumber \\
	& 3 \left( \frac{1}{a_\CD} \frac{{\mathrm d} a_\CD}{{\mathrm d}t} \right)^2 = 
		 \; 8 \pi G \baverage{ \, \epsilont \, } + {\tilde\Lambda}_\CD - \frac{1}{2} \big\langle{\SRt}\big\rangle_\CD
			- \frac{1}{2} \tilde{\CQ}_\CD \; ; \nonumber \\
& \ddt{} \tilde{\CQ}_\CD + 6 H_\CD \tilde{\CQ}_\CD + \ddt{} \big\langle{\SRt}\big\rangle_\CD
			+ 2 H_\CD \,\big\langle{\SRt}\big\rangle_\CD + 4 H_\CD \tilde{\CP}_\CD \nonumber \\
	& {}= 16 \pi G \left( \ddt{} \baverage{\epsilont}
			+ 3 H_\CD \baverage{\epsilont + \tilde p} \right)
			+ 2 \ddt{} {\tilde\Lambda}_\CD \; . 
	\label{foliationdependentequations}
\end{eqnarray}
The first terms on the right-hand side of the last equation also obey an averaged energy balance equation sourced by the non-perfect-fluid parts of the energy-momentum tensor.
We observe a time- and domain-dependent contribution from the cosmological constant,
${\tilde\Lambda}_\CD \equiv \cc \langle N^2 / \gamma^2 \rangle_\CD$, 
and two terms $\tilde{\CQ}_\CD$ and $\tilde{\CP}_\CD$
denoting the intrinsic \textit{kinematical} and \textit{dynamical backreaction} terms, respectively. These are defined in terms of the rescaled fluid variables as follows:
\begin{eqnarray}
	&\tilde{\CQ}_\CD 
		 \equiv \frac{2}{3} \baverage{ \left( \Thetat - \baverage{\Thetat} \right)^2 }
			- 2 \baverage{ \sigmat^2 } + 2 \baverage{ \omegat^2 } \, ;
\nonumber \\ 
	&\tilde{\CP}_\CD 
		 \equiv \baverage{\tilde\CA} + \baverage{\Thetat \frac{\gamma}{N} \ddt{} \left(\frac{N}{\gamma} \right)} \, .
 \label{foliationdependentbackreaction}
\end{eqnarray}
The \textit{dynamical backreaction} thus consists of an acceleration $4-$divergence and of a contribution that captures the \textit{rate of desynchronization of the clocks}, with the proper time $\tau$ of the fluid obeying ${\mathrm d}\tau / {\mathrm d}t = N / \gamma =  M$.
By defining an effective diagonal energy-momentum tensor with the following 
effective sources:\footnote[7]{We have defined
new backreaction variables: $\tilde\CW_\CD$ for the deviation of the averaged fluid $3-$curvature from a 
constant-curvature behaviour, $\tilde\CW_\CD \equiv \langle{\SRt}\rangle_\CD - 6 k_\CD / (a_\CD)^2$,
and $\tilde\CL_\CD$ for the deviation from the cosmological constant $\cc$, $\tilde\CL_\CD \equiv \tilde\Lambda_\CD - \cc$.
$k_\CD$ is an \textit{a priori} domain-dependent arbitrary constant which can be set to
$k_\CD \equiv (a_\CD )^2 (t_{\mathbf i})\langle{\SRt}\rangle_{\CD}(t_{\mathbf i})/6$. In the standard cosmological model it is assumed that the cosmological constant $\cc$ models Dark Energy; the averaged equations show that we then also have to account for \textit{Dark Energy backreaction} $\tilde\CL_\CD$ in cases where $N \ne \gamma$.}
\begin{eqnarray}
 \epsilon_{\rm eff} & \equiv & \baverage{\epsilont} - \frac{\tilde\CQ_\CD}{16 \pi G}  - \frac{\tilde\CW_\CD}{16 \pi G}  + \frac{\tilde\CL_\CD}{8 \pi G}   \, ; 
\nonumber \\
 p_{\rm eff} & \equiv & \baverage{\tilde p} - \frac{ \tilde\CQ_\CD}{16 \pi G} + \frac{ \tilde\CW_\CD}{48 \pi G} - 
 \frac{ \tilde\CL_\CD}{8 \pi G} - \frac{\tilde\CP_\CD}{12 \pi G}  \, , \qquad \qquad
\label{effectivesources1}
\end{eqnarray}
the set of effective cosmological equations can be cast into `Friedmannian form':
\begin{eqnarray}
 3 \, \left( \frac{1}{a_\CD}\frac{{\rm d} a_\CD}{{\rm d}t}\right)^2  =  8 \pi G \,\epsilon_{\rm eff} - 3 \frac{k_\CD}{(a_\CD)^2} + \cc \; ;
 \nonumber \\
 3 \, \frac{1}{a_\CD} \frac{{\rm d}^2 a_\CD}{{\rm d}t^2}  =  - 4 \pi G \,(\epsilon_{\rm eff} + 3 \, p_{\rm eff}) + \cc 
  \label{eq:friedmann_raych} \; ; \nonumber \\
  \ddt{} \epsilon_{\rm eff} + 3 \, H_\CD \, \left( \epsilon_{\rm eff} + p_{\rm eff} \right) =0 \; ,
\label{effectiveequations1}
\end{eqnarray}
where the last equation, the \textit{effective energy conservation law}, is equivalent to the \textit{integrability condition}. The set of equations (\ref{effectiveequations1}) needs a closure condition, e.g. an \textit{effective equation of state} that relates 
$\epsilon_{\rm eff}$, $p_{\rm eff}$ and $a_\CD$.

\section{Effective cosmological equations in the fluid proper time foliation}
\label{intrinsic}

Starting from an arbitrary Cauchy hypersurface, one can globally construct a $3 + 1$ foliation the slices of which are obtained by transporting the initial hypersurface through the (general) $4-$velocity $\bm u$ of the fluid.
Each hypersurface of this foliation corresponds to a constant value of proper time $\tau$, measured along the fluid world lines and being set to $\tau_{\mathbf i} \equiv t_{\mathbf i}$ on the initial slice. The proper time $\tau$ can thus be used to label the hypersurfaces, defining a global time parameter. The same construction can be performed from any choice of the initial Cauchy hypersurface, identifying what we call the class of \textit{fluid proper time foliations}. (See also \cite{Ellisbook}, chapter 4.1.) 

Such a construction sets the normal vector $\bm n$ and the lapse $N$, which in this case equals the Lorentz factor: $N=\gamma$. A natural choice for the shift vector $\bm N$ would be $\bm N = N \bm v$ (for which $N=\gamma$ implies $N^2 - N^\mu N_\mu = 1$), identifying the points on each hypersurface that correspond to the same fluid element. However, the choice of a shift does neither affect the definition of our averaging formalism nor the resulting averaged equations. Apart from the case of irrotational dust, the hypersurfaces of a fluid proper time foliation cannot
be fluid-orthogonal, namely a tilt must be present. As we shall see, this choice
carries a number of advantages in the context of the averaging problem.

Within a fluid proper time foliation, the general \textit{volume expansion} and \textit{acceleration laws} for the fluid scale factor $a_\CD$ (together with their \textit{integrability condition}), (\ref{foliationdependentequations}), reduce to the following effective cosmological equations: 
\begin{eqnarray}
 3 \, \left( \frac{\dot a_\CD}{a_\CD}\right)^2  =  8 \pi G \,\epsilon_{\rm eff} - 3 \frac{k_\CD}{(a_\CD)^2} + \cc \quad ; \quad
3 \frac{\ddot a_\CD}{a_\CD}  =  - 4 \pi G (\epsilon_{\rm eff} + 3 \, p_{\rm eff}) + \cc \; ;
  \label{eq:friedmann_lagrange2} \nonumber \\
{\dot\epsilon}_{\rm eff} + 3 \, H_\CD \, \left( \epsilon_{\rm eff} + p_{\rm eff} \right) = 0 \; .
\end{eqnarray}
The overdot denotes the covariant derivative with respect to proper time. 
The effective energy density $\epsilon_{\rm eff}$ and effective pressure $p_{\rm eff}$,
as defined in (\ref{effectivesources1}), become 
\begin{eqnarray}
\epsilon_{\rm eff} & = & \baverage{\epsilon} - \frac{\QD}{16 \pi G}  - \frac{\WD}{16 \pi G} \, ; 
\label{effectivesources_lagrange1} \nonumber \\
p_{\rm eff} & = & \baverage{p} - \frac{\QD}{16 \pi G} + \frac{\WD}{48 \pi G}  - \frac{\PD}{12 \pi G} \; , \quad \qquad
\label{effectivesources2}
\end{eqnarray}
with $\QD$ as given by (\ref{foliationdependentbackreaction}) with non-rescaled variables (since here $M=1$), and where the dynamical backreaction reduces to $\CP_\CD = \langle\CA\rangle_\CD$, removing the contribution from clock desynchronization.
The cosmological constant deviation $\tilde\CL_\CD$ vanishes, and the curvature deviation term $\tilde\CW_\CD$ reduces to $\WD = \langle{\SR}\rangle_\CD - 6 k_\CD / (a_\CD)^2$. 

We emphasize that the above system and the corresponding proper time foliation choices are 
covariantly defined, i.e. are coordinate-independent \cite{heinesen1}. 
For concrete calculations of local variables, a specific set of coordinates may then be chosen depending on the problem being investigated. For instance, for the formation of structure in relativistic Lagrangian perturbation theory \cite{rza1}, an appropriate set can be constructed as follows.
First, as for the hypersurfaces labeled in terms of proper time, we can introduce \textit{spatial labels} $X^i$ to identify each fluid element in the general threading congruence defined by $\bm u$, which can always be relabeled in this covariant framework. (The spatial labels $X^i$ provide the same identification of points as the shift vector choice $\bm N = N \bm v$.) Second, for any given foliation, these labels may be used as a set of spatial coordinates propagating along the fluid flow lines. These are \textit{comoving} (or \textit{Lagrangian}) spatial coordinates, where the spatial coordinate velocity (hence the spatial components of $u^\mu$) vanish. We name this choice \textit{comoving description} of the fluid, in conformity with the literature. This description is a `weak' form of a \textit{Lagrangian description} of the fluid where in addition $\tau$ is used as the time-coordinate. The coordinate assignment ($X^i , \tau$) provides $u^{\mu} = (1,0,0,0)$. This defines \textit{Lagrangian observers} who in the standard model are called \textit{fundamental observers}.

\section{Conclusion and Discussion}

Looking at the set of equations (\ref{foliationdependentequations}) and their backreaction terms (\ref{foliationdependentbackreaction}) we appreciate that the explicit foliation dependence is solely given in terms of the threading lapse $M=N/\gamma$.
In the \textit{fluid proper time foliation} we have $M= 1$, which does not differ significantly from the value of the threading lapse in the metric form $\lbrace$(\ref{quasinewton1}) and (\ref{quasinewton2})$\rbrace$ for the usual assumptions
$N=1 + \varepsilon \ ; \ |\varepsilon | \ll 1$ and $\gamma = 1 + \zeta \ ; \ \zeta \ll 1$.
The remaining foliation dependence of the amount of backreaction 
arises in the realization of the averaged model, when integration of local variables is performed over specific hypersurfaces that are not fully determined by $N/\gamma$ due to the degeneracy of this ratio.

Let us now narrow down the class of relevant foliations, focussing on matter-dominated model universes. We think of a cosmological coarse-graining that smoothes over scales where vorticity, velocity dispersion and pressure play a role.
In view of observations one can then reasonably assume the existence of a class of foliations where the hypersurfaces reflect statistical homogeneity and isotropy and in which the motions of all coarse-grained fluid elements are non-relativistic, i.e. $\gamma \simeq 1$, thus identifying a class relevant to cosmology (see also the related discussion in \cite{rasanen}). This implies that the tilt is negligible on these scales, $u^{\mu} \simeq n^{\mu}$, and, in view of the negligible pressure gradients over the coarse-graining scale, that the lapse function can be set to $N \simeq 1$.\footnote[8]{Another issue arises if we also consider the coarse-graining of `time' that may accumulate an effective lapse during differing histories of voids and clusters,  \textit{cf.} the `timescape scenario' \cite{wiltshire}.} Overall this estimates $M$ to be close to a Lagrangian description, $M \simeq 1$, while the domain of integration selected by the hypersurfaces is bound to small variations in spacetime, since these hypersurfaces are constrained to remain almost orthogonal to $\bm{u}$ everywhere. Thus, these conditions imply only small variations of the large-scale backreaction terms (of the order of the deviations of the lapse and the Lorentz factor from $1$) under a change of cosmological spacetime foliation. 
Explicit bounds on such variations will be investigated in a forthcoming paper \cite{heinesen2}. These covariantly defined requirements cannot be reproduced in a coordinate-dependent setting such as that used in~\cite{Adamek2017}.
The variations can of course be larger when going beyond this restricted class of foliations that are favoured on cosmological scales, as it would, e.g. be needed for evaluating backreaction on smaller scales. 
These scales, where tilt, vorticity and pressure gradients matter, can be treated as well within the general framework introduced in this letter.

We emphasize that the lapse and the Lorentz factor only depend on the normal vector flow, and not on its derivatives, 
allowing for strong constraints on variations of the backreaction with the foliation when the normal vector itself is constrained. In our formalism, the kinematical backreaction does not involve the extrinsic curvature, which depends on derivatives of the normal vector. It features instead derivatives of the relative velocities of the fluid elements (such as $\Theta$). These foliation-independent scalars can be large despite velocities themselves being small (cf., footnote \ref{footnote2}), allowing for large backreaction.  
We remark in this context that the fact that $M\simeq 1$ in the metric form $\lbrace$(\ref{quasinewton1}) and (\ref{quasinewton2})$\rbrace$, together with the smallness assumptions made, does not mean that the estimates of backreaction in paper \cite{Adamek2017} fall within our conclusions about the small impact of the foliation choice. These authors employ an extrinsic averaging formalism where dependencies on \textit{derivatives} of the normal vector $\bm n$ (and, thus, on derivatives of $N$ and $\gamma$) are introduced in the backreaction terms \textit{via} the dependence on the extrinsic curvature of the hypersurfaces. 
This may lead to unphysical foliation dependence of backreaction, just because the variables to be averaged are defined from the hypersurfaces themselves, i.e. they characterize the properties of a family of extrinsic observers.
(We consider this additional foliation dependence `unphysical', since 
such observers only exist as a mathematical abstraction.)\footnote[9]{We also remark that if backreaction happens to be zero in one foliation (e.g. if subjected to a $3-$torus constraint on a flat space section \cite{newtonianbackreaction}), and if it is represented by a small number in its numerical realization, a still small but nonzero backreaction parameter in another foliation could suggest a \textit{ratio} of several orders of magnitude, even if both estimates were in reality comparable.}

The \textit{fluid proper time foliation} and its representation in terms of the \textit{Lagrangian description} appear to be natural choices for an effective cosmology.
These settings should not be disqualified in favour of a choice of foliation extrinsic to the fluid by emphasizing the need to avoid singularities. For example, evolving a dust matter model implies the development of shell-crossing, as discussed in \cite{rza1}, as a manifestation of the breakdown of the dust approximation. Improving the matter model may or may not avoid these or other (e.g.\ black hole) singularities.

A possible shortcoming of the proper time foliations relates to the spacelike character of the corresponding hypersurfaces generated from the evolution of a single fluid. While such foliations are always well-defined under the assumption that the fluid flow contains no singularity, one has to guarantee that the hypersurfaces, generated from the
initial spacelike slice, remain spacelike for all times considered.
This will hold at least locally in general and globally for an irrotational dust model with a fluid-orthogonal initial hypersurface (since the whole foliation will then be fluid-orthogonal). The construction of a proper time foliation is based on the choice of an initial Cauchy hypersurface, which has to be specified; it may be best anchored to the last scattering surface at the CMB epoch. These aspects have to be judged within specific applications.

The proper time choice can also be criticized because it requires following the details of inhomogeneities developing in the fluid. This latter view originates, however, from looking at a single realization of the fluid's evolution and a single inhomogeneous metric. What the averaged equations embody goes beyond the picture obtained from a single realization of the metric. Changing the metric will change the morphology of the averaging domain and the volume element, but we are entitled to implement the cosmological model through a statistical ensemble of realizations. With this statistical interpretation of the averaged equations, the effective cosmological equations no longer trace individual metric variations as suggested by 
a one-metric-based picture. In this context, a further important question for the definition of statistical hypersurfaces will be whether the tilt, depending on physics on smaller scales, would average out to provide an effective flow-orthogonal foliation on cosmological scales. Follow-up work is dedicated to explicitly implementing these statistical aspects.

\bigskip

\noindent
{\bf Acknowledgments:}
This work is part of a project that has received funding from the European Research Council (ERC) under the European Union's Horizon 2020 research and innovation programme (grant agreement ERC adG No.\ 740021--ARThUs, PI: TB). 
PM is supported by a `sp\'ecifique Normalien' Ph.D. grant; XR is grateful for visiting support and hospitality at CRAL.
We thank L\'eo Brunswic, Mauro Carfora, Asta Heinesen, Nezihe Uzun, and David Wiltshire for valuable discussions,
as well as the anonymous referees for their insightful remarks that helped clarifying the definition of the fluid proper time foliations.
\section*{References}

\end{document}